# Study of fluctuation conductivity in $YBa_2Cu_3O_{7-\delta}$ films in strong magnetic fields

E. V. Petrenko[1], L. V. Omelchenko[1], Yu. A. Kolesnichenko[1], N. V. Shytov[1],
K. Rogacki[2], D. M. Sergeyev[3], and A. L. Solovjov[1,4]

[1]*B. Verkin Institute for Low Temperature Physics and Engineering of NAS of Ukraine, Kharkiv 61103, Ukraine*

[2]*Institute for Low Temperatures and Structure Research, Polish Academy of Sciences, Wroclaw 50-950, Poland*

[3]*K. Zhubanov Aktobe Regional State University, Aktobe 030000, Kazakhstan*

[4]*The Faculty of Physics, V. N. Karazin Kharkiv National University, Kharkiv 61022, Ukraine*
E-mail: petrenko@ilt.kharkov.ua

We report the effect of the *ab*-plane magnetic field $B$ up to 8 T on the resistivity $\rho(T)$ and fluctuation conductivity $\sigma'(T)$ in $YBa_2Cu_3O_{7-\delta}$ thin films. As expected, up to ~ 2.5 T the magnetic field monotonously increases $\rho$, the width of the resistive transition, $\Delta T_c$, and coherence length along the $c$ axis, $\xi_c(0)$, but decreases both $T_c$ and the range of superconducting (SC) fluctuations $\Delta T_{fl}$. The fluctuation conductivity exhibits a crossover at characteristic temperature $T_0$ from the 3D Aslamasov–Larkin (AL) theory near $T_c$ to the 2D fluctuation theory of Maki–Thompson (MT). However, at $B = 3$ T, the MT term is completely suppressed, and above $T_0$ $\sigma'(T)$ is unexpectedly described by the fluctuation contribution of 2D AL, suggesting the formation of a 2D vortex lattice in the film under the action of a magnetic field. At the same time, $\Delta T_{fl}$ sharply increases by a factor of about 7, and $\xi_c(0)$ demonstrates a very unusual dependence on $T_c$ when $B$ increases above 3 T. Our results demonstrate the possibility of the formation of a vortex state in YBCO and its evolution with increasing $B$.

## 1. Introduction

One of the most actual challenges in modern solid-state physics is to develop a theory that can fully describe high-temperature superconductors (HTSCs). Unfortunately, a serious obstacle remains the lack of a clear understanding of the physics of internal interactions in multicomponent compounds such as HTSCs, in particular, the mechanism of superconducting (SC) pairing, which makes it possible to have a superconducting transition temperature $T_c > 100$ K [1, 2]. It is believed that the study of unusual features of the normal state of HTSC, primarily such as the pseudogap (PG), sheds light on the microscopic mechanism of high-temperature superconductivity [3, 4] (and references therein). Among HTSCs, one can distinguish a class of metal oxides with an active plane $CuO_2$ such as, for example, $YBa_2Cu_3O_{7-\delta}$ (or YBCO), called cuprates. In addition to high $T_c$ and PG, these substances have a low density of charge carriers $n_f$, strong electronic correlations, quasi-two-dimensionality and, as a consequence, strong anisotropy of electronic properties [5–8]. In particular, the coherence length along the *ab* plane, which determines the size of Cooper pairs in HTSCs, is $\xi_{ab}(T) \sim 10\xi_c(T)$, where $\xi_c(T)$ is the coherence length along the $c$ axis [3].

It is the low charge carrier density, $n_f$, that promotes the formation of paired fermions in cuprates below the characteristic temperature $T^* \gg T_c$, the so-called local pairs (LPs) [7, 9], which are considered responsible for the formation of the PG ([3, 9, 10] and references therein). At high temperatures $T \leq T^*$ LPs appear in the form of strongly bound bosons (SBBs), which obey the Bose–Einstein condensation theory and can form only in systems with small $n_F$ [10]. SBBs are small, since $\xi_{ab}(T^*) \sim 10$ Å in YBCO, and very tightly bound pairs, since, according to the theory [5, 8], the binding energy in a pair $\varepsilon_b \sim 1/\xi$. As a result, SBBs are not destroyed by thermal and other fluctuations. However, as the temperature decreases, $\xi$ and, consequently, the size of the pairs grows, and the SBBs are gradually transformed into fluctuating Cooper pairs (FCPs), which already obey the BCS theory near $T_c$ [9–11]. We emphasize that near $T_c$, when $\xi_c(T)$ exceeds the size of the YBCO unit cell along the $c$ axis: $\xi_c(T) > d$, the quasi-two-dimensional state of HTSCs changes to the 3D state [3, 11], since the condensation of FCPs into the SC state is possible only from the 3D state [5, 8].

We also note that at $T \leq T^*$ not only the HTSC resistance deviates downward from the linear dependence at high temperatures, but also the density of states at the Fermi level begins to decrease [12, 13], which by definition is called a pseudogap [4]. At present, it can be considered quite established [14, 15] that at $T \leq T^*$ the Fermi surface rearrangement also begins from a single large hole surface above quantum critical point (QCP) $p^* \sim 0.19$ in YBCO, where the non-superconducting ground state is a correlated metal, to small electron pockets between $p = 0.08$ and 0.16, where charge order prevails at low temperature and finally to small nodal hole pockets below $p = 0.08$, where magnetic order prevails at low temperature [15]. Recall that the hall concentration $p = 0.16$ corresponds to critical doping, whereas $p^* = 0.19$ corresponds to pseudogap critical point (or QCP) [14, 15]. Moreover, it is currently believed to be well established that, at least in Bi2201, below $T^*$, the FS is not a closed surface, but decomposes into separate Fermi arcs [16, 17].

It is believed that a correct understanding of such an unusual phenomenon as the PG state in HTSCs should also answer the question of the mechanism of superconducting pairing in cuprates. To explain the PG state, spin fluctuations [18], charge [15, 19] and spin [14–16] density waves, charge ordering ([15, 16] and references therein) are proposed. However, despite the fact that the interest in the PG study has noticeably increased in recent years [20–23], the physics of the PG state is still not completely clear. At the same time, although the number of works devoted to the study of HTSCs and, in particular, the PG, is extremely large, there is a lack of papers studying the influence of a magnetic field on excess conductivity and especially on fluctuation conductivity (FLC) and PG in cuprates. However, it is the study of the effect of the magnetic field on FLC and PG that can answer the question of which of the mechanisms considered above actually take place in cuprates.

A few attempts have been made to analyze the excess conductivity in YBCO in strong magnetic fields [24–26]. In [24], both $\mathbf{B}\|ab$ and $\mathbf{B}\|c$ orientations of the external magnetic field were used to study FLC in aluminum-doped YBCO single crystals with a system of unidirectional twin boundaries (TBs). Obviously, in this case, the results are deeply affected by the aluminum impurities and TBs. Besides, the applied field does not exceed 1.27 T. In [25], magnetic field $B = 12$ T was applied to study FLC in YBCO composites, which again were doped with Ag, which deeply affected the results. It is not known whether these were films, single crystals, or polycrystals. The authors did not show the evolution of the FLC with a magnetic field, but show only the data at $B = 0$ and 12 T. In [26], the magnetic susceptibility and FLC were measured in an optimally doped ($T_c = 91.1$ K) Y-123 film in fields up to 9 T perpendicular to the *ab* plane. But, surprisingly, as in [25], the authors also did not show the evolution of the FLC with a magnetic field. In addition, no attempts were made to use the Aslamasov–Larkin and Maki–Thompson fluctuation theories to describe FLC as a function of *B*. As a result, the mechanism of the magnetic field influence on the FLC and the vortex motion in YBCO is still unclear.

In this paper, we report on the study of the effect of a magnetic field in the *ab* plane on the resistivity $\rho(T)$ and fluctuation conductivity $\sigma'(T)$ of a $YBa_2Cu_3O_{7-\delta}$ (YBCO) thin film with increasing magnetic field from $B = 0$ to 8 T. Since the magnetic field does not affect the normal-state resistivity, the studies were carried out in the temperature range corresponding to the transition of the film to the SC state. The study can become a key to understanding the formation of a vortex state in YBCO and its evolution with increasing *B*. It is shown that up to ~ 2.5 T magnetic field monotonously increases $\rho$, the width of the resistive transition, $\Delta T_c$, and $\xi_c(0)$, but decreases both $T_c$ and the range of SC fluctuations $\Delta T_{fl}$. At $B = 0$, $\sigma'(T)$ near $T_c$, as expected, is described by the 3D Aslamasov–Larkin (AL) theory and by the 2D fluctuation theory of Maki–Thompson (MT) above the crossover temperature $T_0$. But at $B = 3$ T, the MT term is completely suppressed, and above $T_0$, $\sigma'(T)$ is unexpectedly described by the 2D AL fluctuation contribution. At the same time, $\Delta T_{fl}$ sharply increases by a factor of about 7, and $\xi_c(0)$ demonstrates very unusual dependence on $T_c$ when $B$ increases above 3 T. A detailed discussion of the results obtained is given below.

## 2. Experiment

The epitaxial YBCO films were deposited at $T = 770$ °C in 3 mbar oxygen pressure at $(LaAlO_3)_{0.3}(Sr_2TaAlO_6)_{0.7}$ substrates, as described in [27]. The thickness of the deposited films, $d \sim 100$ nm, was controlled by the deposition time of the respective targets. X-ray analyses have shown that all samples are excellent films with the *c* axis perfectly oriented perpendicular to the $CuO_2$ planes. Next, the films were lithographically patterned and chemically etched into well-defined 2.35×1.24 mm Hall-bar structures. To perform contacts, golden wires were glued to the structure pads using silver epoxy. Contact resistance below 1 Ω was obtained. The main measurements included a fully computerized setup, the Quantum Design Physical Property Measurement System (PPMS-9), using an excitation current of ~ 100 μA at 19 Hz. The four-point probe technique was used to measure the in-plane resistivity $\rho_{ab}(T) = \rho(T)$.

## 3. Results and discussion

### 3.1. Resistivity

The temperature dependence of the resistivity $\rho(T) = \rho_{ab}(T)$ of the $YBa_2Cu_3O_{7-\delta}$ film in the absence of an external magnetic field is shown in Fig. 1. For temperature above

*E. V. Petrenko, L. V. Omelchenko, Yu. A. Kolesnichenko, N. V. Shytov, K. Rogacki, D. M. Sergeyev, and A. L. Solovjov*

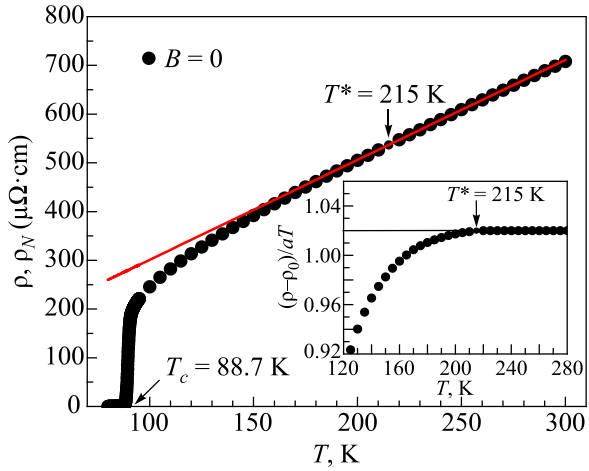

*Fig. 1.* Dependence $\rho(T)$ for YBa$_2$Cu$_3$O$_{7-\delta}$ film in the absence of an external magnetic field ($B = 0$, circles). The solid line determines $\rho_N(T)$ extrapolated to the low-temperature region. The smaller dot corresponds to the $T^*$ temperature. Inset: The method of determining $T^*$, using criterion $[\rho(T) - \rho_0]/aT = 1$ [1].

$T^* = 215$ K and up to 300 K, the $\rho(T)$ dependence is linear and is described by a slope $a = d\rho/dT = 2.050$ μΩ·cm/K. The slope was calculated by approximating the experimentally derived curves and confirmed the linear behavior of $\rho(T)$ with a mean-root-square error of $0.009 \pm 0.002$ in the specified temperature range. The temperature $T^* \gg T_c$ was defined as a temperature at which the resistive curve deviates downward from the linearity (Fig. 1). The more precise approach to determine $T^*$ with accuracy $\pm 1$ K is to explore the criterion $[\rho(T) - \rho_0]/aT = 1$ [1] (inset in Fig. 1), where, as before, $a$ designates the slope of the extrapolated normal-state resistivity, $\rho_N(T)$, and $\rho_0$ is its intercept with the $y$ axis. Both methods give the same $T^* = 215$ K, that is typical for the well-structured YBCO films with $T_c \sim 88$ K and is in good agreement with the literature data [11, 28].

The influence of the magnetic field from $B = 0$ to 8 T on the temperature dependence of $\rho(T)$ is shown in Fig. 2. As can be seen, the magnetic field noticeably broadens the resistive transition and reduces $T_c$, but, as usual, does not affect the normal state of the sample [29, 30]. The straight lines $0.9\rho_n$ and $0.1\rho_n$ allow us to determine both $T_c^{onset}(B)$ and $T_c^{offset}(B)$, respectively. Here, $\rho_n$ is a normal state resistivity in the vicinity of the SC transition.

Measurements of $T_c^{onset}(B)$ make it possible to determine the temperature dependence of the upper critical field $B_{c2}(T)$ (Fig. 3) [31]. One more important curve on the $B$–$T$ phase diagram of HTSCs (Fig. 3) is the irreversibility line [32] that separates the vortex glassy state from the vortex liquid state. This line is determined by the temperature dependence of the irreversibility magnetic fields $B^*(T)$, above which the magnetization curve is reversible [33] and the flowing current forces the vortices to move. It means that energy dissipation appears and the supercurrent vanishes. From this point of view, the irreversibility field of HTCSs plays a similar role as the upper critical field in classical superconductors [32, 33]. Accordingly, $T_c^{offset}(B)$ measurements give $B^*(T)$. Both quantities are shown in Fig. 3 as functions of temperature. Recall that the magnetic field is oriented perpendicular to the $c$ axis of the sample ($\mathbf{B} \| ab$). Straight lines with arrows on the ends show the difference between $T_c^{onset}(B)$ and $T_c^{offset}(B)$, marked as $\Delta T_c(B)$.

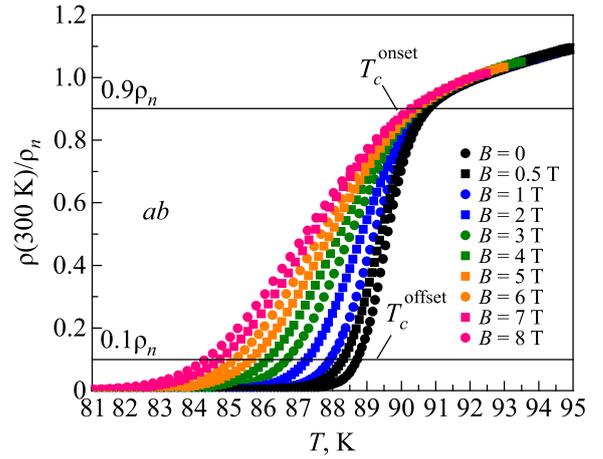

*Fig. 2.* (Color online) Temperature dependences of the resistivity of the YBa$_2$Cu$_3$O$_{7-\delta}$ film measured in the superconducting transition region in a magnetic field 0–8 T parallel to the $ab$ plane: $\rho_n = 203$ μΩ·cm at $T = 92.3$ K is a normal state resistivity in the vicinity of the SC transition. $0.9\rho_n$ and $0.1\rho_n$ determine $T_c^{onset}(B)$ and $T_c^{offset}(B)$, respectively.

### 3.2. Fluctuation conductivity

At high temperatures $T > T^*$, in the resistive measurements, the dependence of the resistivity $\rho_{ab}(T) = \rho(T)$ in HTSCs is linear. Below $T^*$, a deviation of $\rho(T)$ from a linear dependence toward smaller values occurs (refer to Fig. 1), which results in the excess conductivity expressed by $\sigma'(T) = \sigma(T) - \sigma_N(T) = [1/\rho(T)] - [1/\rho_N(T)]$, or

$$\sigma'(T) = \frac{\rho_N(T) - \rho(T)}{\rho(T)\rho_N(T)}, \quad (1)$$

where $\rho_N(T) = aT + b$ is the sample resistivity in the normal state extrapolated to the low-temperature region [11, 28, 34, 35]. It should be noted that, according to the model [18], the linear dependence of $\rho(T)$ above $T^*$ is the normal state of HTSCs that characterizes by the stability of the Fermi surface [14–16, 18].

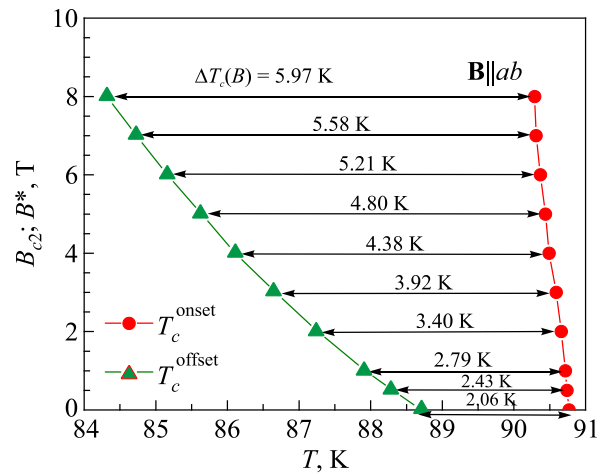

*Fig. 3.* Upper critical field $B_{c2}(T)$ (criterion $0.9\rho_n$, ●) and irreversibility field $B^*$ (criterion $0.1\rho_n$, ▲) as a function of temperature for studied YBa$_2$Cu$_3$O$_{7-\delta}$ film. Magnetic field is oriented in $ab$ plane. Straight lines with arrows on the ends show the difference between $T_c^{onset}(B)$ and $T_c^{offset}(B)$, marked as $\Delta T_c(B)$.

According to recent concepts [3, 9, 10, 34–39], a small value of the coherence length in conjunction with a quasi-layered structure of the HTSCs leads to the formation of a noticeable area of superconducting fluctuations on $\rho(T)$ in the vicinity of $T_c$, where $\sigma'(T)$ follows conventional fluctuation theories [3, 11, 40–42]. At the same time, changes in oxygen content, the presence of impurities and/or structural defects have a considerable impact on $\sigma'(T)$ and, accordingly, on the implementation of various FLC modes above $T_c$ [11, 43–45].

The fluctuation conductivity of the YBCO film under study was determined by analyzing the excess conductivity, which was calculated by the standard method according to formula (1). The FLC analysis was performed within the model of local pairs, in which the presence of paired fermions (LPs) in HTSCs is assumed in the temperature range $T_c < T < T^*$ [3, 9, 10, 23, 36]. First, the mean-field temperature $T_c^{mf} > T_c$ limiting the region of critical fluctuations near $T_c$, where the mean-field theory does not work, has to be determined [46]. In addition, $T_c^{mf}$ determines the reduced temperature

$$\varepsilon = \frac{T - T_c^{mf}}{T_c^{mf}}, \quad (2)$$

which is used in all equations. From this, it is clear that the correct determination of $T_c^{mf}$ plays a key role in the calculations of FLC. At the vicinity of $T_c$, the coherence length in the $c$ axis $\xi_c(T)$ is greater than $d$. Here $d \approx 11.7$ Å [47] is the $c$ axis lattice parameter of the YBCO unit cell [42, 48]. In this case, the fluctuating Cooper pairs are combined throughout the superconductor and form a three-dimensional (3D) state of the HTSC [11, 42, 48]. Therefore, at the proximity of $T_c$, the FLC can be described by the 3D equation of the Aslamazov–Larkin theory [49, 50] with the critical exponent $\lambda = -1/2$, which determines the FLC in any 3D system:

$$\sigma'_{3DAL}(T) = C_{3D} \frac{e^2}{32\hbar\xi_c(0)} \varepsilon^{-1/2}. \quad (3)$$

Here $\sigma'(T) \sim \varepsilon^{-1/2}$. Simple algebra yilds $\sigma'^{-2}(T) \sim \varepsilon \sim T - T_c^{mf}$, which vanishes at $T = T_c^{mf}$ (see Fig. 4) and thus allows determining both $T_c^{mf}$ and $\varepsilon$ with high accuracy [11, 29, 49, 51]. Also in Fig. 4, the arrows show $T_c$ and the Ginzburg temperature $T_G$, down to which the fluctuation theories are valid [42, 48]. The temperature of the 3D–2D crossover $T_0$ limits the area of 3D fluctuations. Notably, above $T_0 = 90.89$ K (refer to Fig. 4), the data deviate to the right from the linear dependence, indicating the presence of 2D Maki–Thompson contribution to the FLC [42, 48]. Having determined $\varepsilon$, we construct the dependences $\ln \sigma'(\ln \varepsilon)$ at different $B$ (Figs. 5 and 6). Upper panel of Fig. 5 shows the corresponding dependence for the case of abcense of magnetic field. As expected, at the vicinity of $T_c$, in the interval $T_G - T_0$ ($\ln \varepsilon_0 = -4.28$), the FLC is well modelled by the 3D AL fluctuation contribution (3). In double logarithmic coordinates, this is the 3D AL line with a slope $\lambda = -1/2$. As mentioned above, it implies that a classical three-dimensional FLC materializes in an HTSC when $T \to T_c$ and $\xi_c(T) > d$ [35, 39, 49]. Above the crossover temperature $T_0$, $\xi_c(T) < d$ [39, 42, 48, 49], and this is no longer a 3D regime. However, as before, $\xi_c(T) > d_{01}$, where $d_{01} \approx 3.5$ Å is the separation of the conducting planes of $CuO_2$ in YBCO [47]. Thus, up to temperature $T_{01}$ ($\ln \varepsilon_{01} = -2.16$, Fig. 5, upper panel) $\xi_c(T)$ connects the inner planes of $CuO_2$ through the Josephson interaction [39, 42]. This is the 2D FLC regime, which is perfectly approximated by the 2D MT equation (curve 2D MT) of the Hikami–Larkin theory for HTSCs [48]:

$$\sigma'_{2DMT}(T) = C_{2D} \frac{e^2}{8d\hbar} \frac{1}{1 - \alpha/\delta} \ln\left(\frac{\delta}{\alpha} \frac{1 + \alpha + \sqrt{1 + 2\alpha}}{1 + \delta + \sqrt{1 + 2\delta}}\right) \varepsilon^{-1}. \quad (4)$$

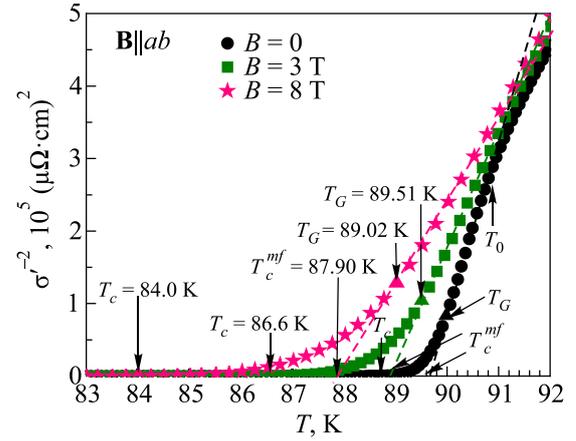

*Fig. 4.* Dependences of $\sigma'^{-2}(T)$ for the $YBa_2Cu_3O_{7-\delta}$ film at $B = 0$ (●); 3 (■) and 8 (★) T. Arrows indicate $T_c$, $T_c^{mf}$, the Ginzburg temperature $T_G$, marked by ▲, and the crossover temperature $T_0$. Straight dashed lines designate the linear approximations of $\sigma'^{-2}(T)$.

Above $T_{01}$, the experimental points finally deviate downward from the theory (Fig. 5) implying that the classical fluctuation theories are no longer valid. Thus, $T_{01}$ limits the region of SC fluctuations $\Delta T_{fl} = T_{01} - T_G$ from above. Conversely, $T_G$ limits the region of SC fluctuations

*E. V. Petrenko, L. V. Omelchenko, Yu. A. Kolesnichenko, N. V. Shytov, K. Rogacki, D. M. Sergeyev, and A. L. Solovjov*

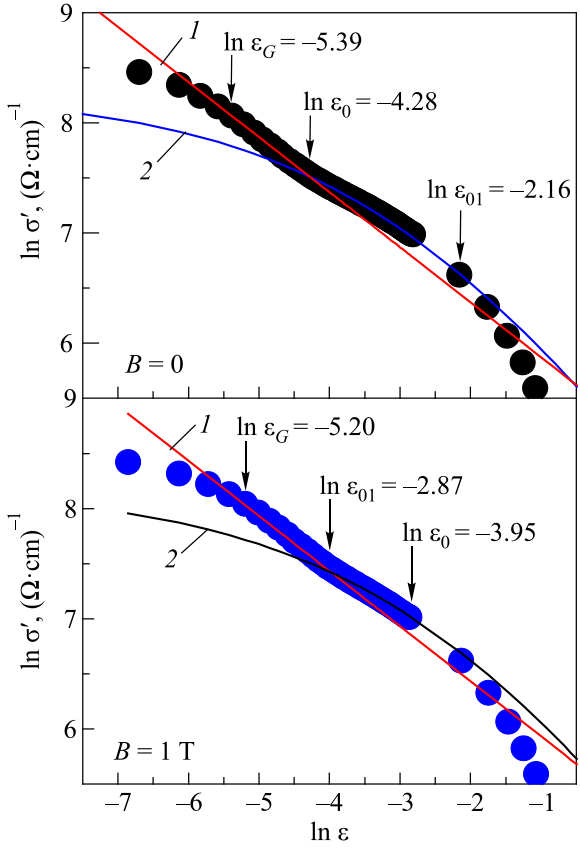
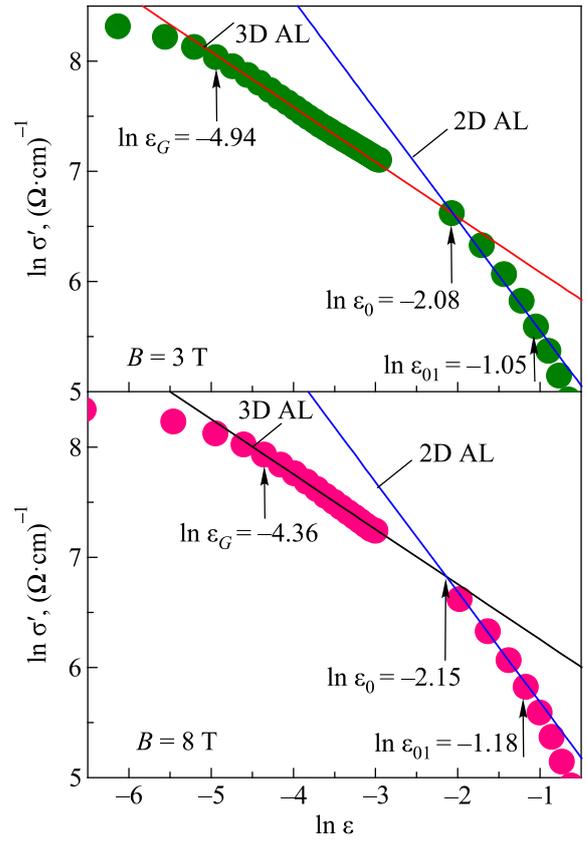

*Fig. 5.* Upper panel: $\ln \sigma'$ vs $\ln \varepsilon$ for the YBCO at $B = 0$ in comparison with fluctuation theories: 3D AL (line *1*) and 2D MT (curve *2*). The $T_{01}$ ($\ln \varepsilon_{01}$) determines the range of SC fluctuations, $T_0$ ($\ln \varepsilon_0$) is the temperature of the 3D–2D crossover and $T_G$ ($\ln \varepsilon_G$) is the Ginzburg temperature. Lower panel: The same dependences for $B = 1$ T.

*Fig. 6.* The same dependences as in Fig. 5 for $B = 3$ and 8 T. In both cases, MT contribution is completely suppressed and $\sigma'$ above $T_0$ is described by 2D AL (blue line). As before, the region of SC fluctuations is limited from above by $T_{01}$.

from below. As a result, below $T_G$, designated as $\ln \varepsilon_G$ in Figs. 5 and 6, the experimental points also deviate downward from the theory, suggesting the transition to the range of critical fluctuations or fluctuations of the SC order parameter $\Delta$ just near $T_c$, where $\Delta < kT$ [46, 49].

At $T_0$, $\xi_c(T_0) = d = 11.7$ Å, which allows us to determine $\xi_c(0)$ [11, 39, 42, 51]

$$\xi_c(0) = d\sqrt{\varepsilon_0}. \quad (5)$$

Taking into account that $\ln \varepsilon_0 = -4.28$ (Fig. 5, upper panel) and using Eq. (5), we get $\xi_c(0) = (1.38 \pm 0.02)$ Å ($B = 0$), which is in a good agreement with $\xi_c(0) = (1.65 \pm 0.02)$ Å obtained for well-structured YBCO film with a slightly lover $T_c = 87.4$ K (sample F1 in [11]), but almost 1.6 times the coherence length along the *c* axis obtained for an optimally doped untwined YBCO single crystal with $T_c = 91.6$ K (sample A1 in [52]). Besides, in both films the region of SC fluctuations is approximately the same: $\Delta T_{\text{fl}} = T_{01} - T_G = 100$ K $- 90.06$ K $\approx 9.9$ K (Table 1). Both findings indicate the better agreement between the properties of films and suggest the validity of classical relation [46]:

$$\xi_0 \sim \frac{\hbar v_F}{\pi \Delta(0)} \quad (6)$$

taking into account the fact that superconducting gap $\Delta(0) \sim k_B T_c$ [46] and in YBCO films the Fermi velocity $v_F$ is almost independent on $T_c$ [11]. In addition, evidently that $\xi_c(T_{01}) = d_{01}$, and, since $\xi_c(0)$ has already been defined by Eq. (5), we can calculate $d_{01}$ from the relation $\xi_c(0) = d\sqrt{\varepsilon_0} = d_{01}\sqrt{\varepsilon_{01}}$. For $B = 0$, calculations give $d_{01} = (4.1 \pm 0.2)$ Å, in good agreement with the literature data [51, 52].

In the above equations

$$\alpha = 2\left(\frac{\xi_c(0)}{d}\right)^2 \varepsilon^{-1} \quad (7)$$

is a coupling parameter;

$$\delta = 1.203 \frac{l}{\xi_{ab}} \frac{16}{\pi \hbar} \left[\frac{\xi_c(0)}{d}\right]^2 k_B T t_\varphi \quad (8)$$

is the pair-breaking parameter, and the phase relaxation time $\tau_\varphi$ is determined by the equation

$$\tau_\varphi \beta T = \pi \hbar / 8 k B \varepsilon = A / \varepsilon, \quad (9)$$

*Study of fluctuation conductivity in* YBa$_2$Cu$_3$O$_{7-\delta}$ *films in strong magnetic fields*

Table 1. Parameters of FLC analysis for YBa$_2$Cu$_3$O$_{7-\delta}$ film depending on the applied magnetic field

| $B$, T | $T_c$, K | $T_c^{mf}$, K | $T_0$, K | $\ln \varepsilon_0$ | $T_{01}$, K | $\ln \varepsilon_{01}$ | $T_G$, K | $\ln \varepsilon_G$ | $\Delta(T_{fl})$, K | $\xi_c(0)$, Å | $\xi_c(0)_{theory}$, Å |
|---|---|---|---|---|---|---|---|---|---|---|---|
| 0 | 88.7 | 89.65 | 90.89 | –4.28 | 100.0 | –2.16 | 90.06 | –5.39 | 9.9 | 1.38 | 1.38 |
| 0.5 | 88.3 | 89.50 | 91.14 | –4.00 | 96.8 | –2.50 | 89.89 | –5.43 | 6.9 | 1.58 | 1.39 |
| 1 | 87.9 | 89.40 | 91.12 | –3.95 | 94.5 | –2.87 | 89.89 | –5.20 | 4.6 | 1.62 | 1.39 |
| 2 | 87.2 | 89.06 | 92.69 | –3.20 | 94.0 | –2.89 | 89.60 | –5.10 | 4.4 | 2.36 | 1.40 |
| 3 | 86.6 | 88.88 | 100 | –2.08 | 120.0 | –1.05 | 89.51 | –4.94 | 30.5 | 4.13 | 1.41 |
| 4 | 86.0 | 88.67 | 100 | –2.06 | 120.0 | –1.04 | 89.44 | –4.75 | 30.6 | 4.18 | 1.42 |
| 5 | 85.6 | 88.46 | 100 | –2.04 | 115.0 | –1.20 | 89.24 | –4.73 | 25.8 | 4.22 | 1.43 |
| 6 | 84.9 | 88.30 | 100 | –2.02 | 115.0 | –1.20 | 89.26 | –4.51 | 25.8 | 4.26 | 1.44 |
| 7 | 84.5 | 88.09 | 98.4 | –2.15 | 115.0 | –1.19 | 89.10 | –4.47 | 25.9 | 3.99 | 1.45 |
| 8 | 84.0 | 87.90 | 98.1 | –2.15 | 115.0 | –1.18 | 89.02 | –4.36 | 26.0 | 3.99 | 1.46 |

where $A = 2.998 \cdot 10^{-12}$ s·K. Here the factor $\beta = 1.203(l/\xi_{ab})$ with $l$ being the mean free path and $\xi_{ab}$ is the coherence length along the *ab* plane, takes into account the approximation of the clean limit ($l > \xi$), which constantly happens in HTSC because of the smallness of $\xi(T)$ [40–42, 48].

Figures 5 and 6 show the dependences $\ln \sigma'(\ln \varepsilon)$ for magnetic fields of 0, 1, 3, and 8 T. It can be seen that with an increase in $B$, the contribution of 2D MT fluctuations is gradually suppressed, and $\Delta T_{fl} = T_{01} - T_G$ also gradually decreases but only up to ~ 2.5 T (Table 1). Ultimately, somewhat unexpectedly, above $B \approx 3$ T, $\ln \sigma'(\ln \varepsilon)$ at $T > T_0$ is now well described by the 2D AL fluctuation term [50]

$$\sigma'_{2DAL} = C_{2D} \frac{e^2}{16\hbar d} \varepsilon^{-1}, \qquad (10)$$

where in this case $d$ is the thickness of the sample. Simultaneously, both $T_{01}$ ($\ln \varepsilon_{01}$ in the figures), that is, the range of SC fluctuations, and $T_0$ ($\ln \varepsilon_0$ in the figures), that is, $\xi_c(0)$ (Table 1), increase significantly. Indeed, the coherence length $\xi_c(0)$ increases from 1.38 Å ($B = 0$) to 3.99 Å ($B = 8$ T) (Table 1) in agreement with Eq. (6), since $\xi_c(0) \sim 1/T_c$. However, $\xi_c(0)$ measured in a magnetic field demonstrates a rather unusual $\xi_c(0)(T_c)$ (Fig. 7, curve *1*) and is almost 3 times larger than $\xi_c(0)$ vs $T_c$ (Fig. 7, curve *2*) calculated by Eq. (6) at $B = 0$ with $v_F = 1.16 \cdot 10^5$ m/s and proportionality coefficient $K = 0.11$ obtained for F1 in [11] (the last column in Table 1). Since formula (6) is rather simple, one may conclude that magnetic field should somehow increase $v_F$. Interestingly, at $B = 3$ T, $\Delta T_{fl}$ sharply increases by a factor of about 7, then unexpectedly decreases by ~ 1.2 at $B = 5$ T, and then remains independent of $B$ (Table 1). Accordingly, $\xi_c(0)$ also sharply increases at about of 1.75 at $B = 3$ T, then remains almost independent of $B$ and even decreases at $B > 6$ T (Fig. 7).

According to our knowledge, such an unusual transformation of FLC under external influence is observed for the first time. Really, in YBCO films with defects [29, 53] $\ln \sigma'(\ln \varepsilon)$ is always described by the Lawrence–Doniach model [40]:

$$\sigma'_{LD} = C_{LD} \frac{e^2}{16\hbar d} \frac{1}{\varepsilon \sqrt{1 + 2\alpha}}, \qquad (11)$$

which reduces to 3D AL when $T \to T_c$ and to 2D AL at high temperatures [29].

In Y$_{1-x}$Pr$_x$Ba$_2$Cu$_3$O$_{7-\delta}$ single crystals [54], the increase in the number of defects generated by Pr results in both a noticeable decrease in $T_c$ and a large increase in the resistivity. The same result is obtained when YBCO single crystals are irradiated with high-energy electrons [55]. Moreover, in all cases when there is no magnetic field, the width of the resistive transition $\Delta T_c$ remains rather narrow, and the 2D AL behavior of $\ln \sigma'(\ln \varepsilon)$ has never been observed. In addition, magnetic field does not affect the normal-state resistivity above $T_c^{onset}$ [29, 30], as mentioned above. It is also worth noting the absence of visible features on the resistivity curves (Fig. 2) obtained at different values of $B$ in the range of SC fluctuations.

Taking into account all of the above, we can conclude that the influence of an external magnetic field does not reduce to a simple introducing additional disorder into the system. And, in fact, the revealed transformation of FLC

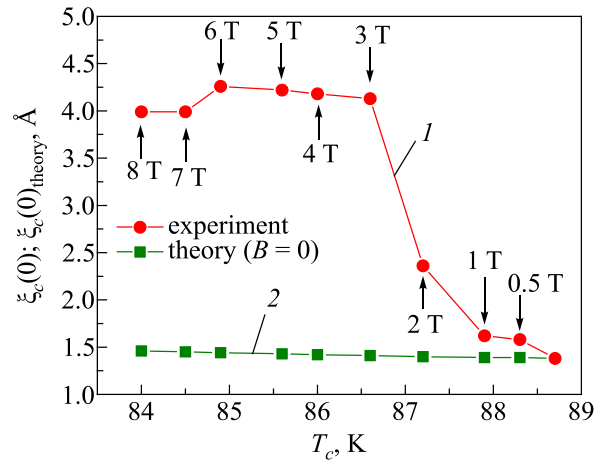

*Fig. 7.* Temperature dependences of $\xi_c(0)$ of the YBa$_2$Cu$_3$O$_{7-\delta}$ film (dots with corresponding fields marked by the arrows) and $\xi_c(0)_{theory}$ (square) calculated at $B = 0$ using Eq. (6).

Low Temperature Physics/Fizika Nizkikh Temperatur, 2021, vol. 47, No. 12        1153

*E. V. Petrenko, L. V. Omelchenko, Yu. A. Kolesnichenko, N. V. Shytov, K. Rogacki, D. M. Sergeyev, and A. L. Solovjov*

from 2D MT to 2D AL under the influence of a magnetic field remains an open question. Most likely, the magnetic field forms a two-dimensional vortex lattice in the film, which leads to the observation of 2D AL FLC above $T_0$, when $B$ exceeds 3 T.

To obtain more detailed information, it is necessary to study additionally the magnetoconductivity $\Delta\sigma_H$. It is known from the literature [56] that the orbital motion of superconducting carriers is strongly suppressed when using the scenario **B**||*ab* instead of **B**||*c*. For this reason, only Zeeman terms contribute to the "magnetic" fluctuation conductivity $\Delta\sigma_H$ in this case. Corresponding experiments are ongoing. Our next article will be devoted to a detailed analysis of $\Delta\sigma_H$, which, according to the relevant literature, is practically not studied due to the complex shape of the ALZ and MTZ contributions.

## Conclusions

We report two-dimensional vortex lattice under influence of the external magnetic field up to 8 T in a YBa$_2$Cu$_3$O$_{7-\delta}$ thin film. The measured fluctuation conductivity exhibits a crossover at a critical field $B \sim 3$ T from a semiclassical weak-field 2D MT dependence to the unexpected high-field 2D AL dependence. In this case, $T_{01}$ sharply increases, and the range of SC fluctuations $\Delta T_{\text{fl}}$ also increases by about 7 times. Accordingly, $\xi_c(0)$ demonstrates a rather unusual dependence on $T_c$ in comparison with the classical $\xi_c(0)$ vs $T_c$, usually observed in YBCO films at $B = 0$ T. The results obtained demonstrate the possibility of the formation of a rather unusual two-dimensional vortex lattice in a YBCO film under the action of in-plane magnetic field, but only when the field exceeds 3 T.

## Acknowledgments

This research was funded in part by the Science Committee of the Ministry of Education and Science of the Republic of Kazakhstan (Grant No. AP08052562).

___________________________

Вивчення флуктуаційної провідності в плівках $YBa_2Cu_3O_{7-\delta}$ в сильних магнітних полях

E. V. Petrenko, L. V. Omelchenko, Yu. A. Kolesnichenko, N. V. Shytov, K. Rogacki, D. M. Sergeyev, A. L. Solovjov

Вивчено вплив магнітного поля в орієнтації *ab*-площини до 8 Тл на питомий опір ρ(*T*) та флуктуаційну провідність σ′(*T*) у тонких плівках $YBa_2Cu_3O_{7-\delta}$. Як очікувалось, до ~ 2,5 Тл магнітне поле монотонно збільшує ρ, ширину резистивного переходу Δ$T_c$ та довжину когерентності вдовж осі *c*, $\xi_c(0)$, але зменшує як $T_c$, так й інтервал надпровідних флуктуацій Δ$T_{fl}$. Флуктуаційна провідність демонструє кросовер при характерній температурі $T_0$ з 3D-теорії Асламазова–Ларкіна (АЛ) поблизу $T_c$ на 2D-флуктуаційну теорію Макі–Томпсона (МТ). Однак при *B* = 3 Тл внесок МТ повністю пригнічується, і вище $T_0$ σ′(*T*) несподівано описується флуктуаційним 2D-внеском АЛ, що свідчить про утворення 2D-вихрової гратки в плівці під дією магнітного поля. У той же час Δ$T_{fl}$ різко зростає приблизно в 7 разів, а $\xi_c(0)$ демонструє дуже незвичну залежність від $T_c$, коли *B* зростає вище 3 T. Отримані результати демонструють можливість утворення вихрового стану в YBCO та його еволюцію зі збільшенням *B*.

Ключові слова: високотемпературні надпровідники, плівки YBCO, надлишкова провідність, флуктуаційна провідність, магнітне поле, довжина когерентності.